\documentstyle[12pt,epsfig]{article}

\def\sss{\scriptscriptstyle}
\def\barp{{\raise.35ex\hbox{${\sss (}$}}---{\raise.35ex\hbox{${\sss )}$}}}
\def\bdbarp{\hbox{$B_d$\kern-1.4em\raise1.4ex\hbox{\barp}}}
\def\bsbarp{\hbox{$B_s$\kern-1.4em\raise1.4ex\hbox{\barp}}}
\def\dbarp{\hbox{$D$\kern-1.1em\raise1.4ex\hbox{\barp}}}

\newcommand{\beq}{\begin{equation}}
\newcommand{\eeq}{\end{equation}}
\newcommand{\beqa}{\begin{eqnarray}}
\newcommand{\eeqa}{\end{eqnarray}}
\def\g{\gamma}
\def\ra{\rightarrow}

\newcommand{\mh}{\mbox{$M_H$}}

\def\journal#1#2#3#4{{\it #1} {\bf #2} (#3) #4}
\def\epj{Euro. Phys. Jour.}
\def\prl{Phys. Rev. Lett.} 
\def\pl{Phys. Lett.}
\def\np{Nucl. Phys.}
\def\ptp{Prog. Theor. Phys.}
  
\def\zp{Z. Phys.}
\def\pr{Phys. Rev.}

\def\mc{{\hat{m_c}}}

\def\o{{\cal O}}
\def\c{C}
\def\cs{{\c_7}} 
\def\cn{{\c_9}}
\def\ct{{\c_{10}}}   
\def\cne{\cn^{\rm eff}}
\def\cse{\cs^{\rm eff}}
\def\m{{\cal M}}
\def\gl{\Gamma} 
\def\g{\gamma}  
\def\l{\ell}  
   
\def\lb{\bar{\l}}

\def\b{{\cal B}}

\def\acp{{{\cal A}_{\rm CP}}}

\def\he{{\cal H}_{\rm eff}}
\def\he{{\cal H}_{\rm eff}}

\def\mh{\hat{m}}
\def\mbh{\mh_b}

\def\mvh{\mh_{K^*}}

\def\qh{\hat{q}}

\def\sh{\hat{s}}

\def\a{{\cal A}}

\def\uh{{\hat{u}}}

\def\be{\begin{equation}}
\def\ee{\end{equation}}
\def\ba{\begin{eqnarray}}
\def\ea{\end{eqnarray}}

\def\be{\begin{equation}}
\def\ee{\end{equation}}
\def\g{\gamma}

\def\mc{m_c}

\def\bea{\begin{eqnarray}}
\def\eea{\end{eqnarray}}
\def\be{\begin{equation}}
\def\ee{\end{equation}}
\def\a{\alpha}
\def\b{\beta}
\def\g{\gamma}

\def\l{\lambda} 
\def\m{\mu}

\newcommand{\BGAMAXS}{B \ra X _{s} + \gamma}
\newcommand{\BGAMAXD}{B \ra X _{d} + \gamma}

\newcommand{\BBGAMAXD}{{\cal B}(B \ra  X _{d} + \gamma)}

\def\beq{\begin{equation}}
\def\eeq{\end{equation}}

\def\Vtdabs{\vert V_{td} \vert}

\def\Vtsabs{\vert V_{ts} \vert}

\newcommand{\go}[1]{\gamma^{#1}}
\newcommand{\gu}[1]{\gamma_{#1}}

\def\sh{\hat s}
\def\mh{\hat m}
\def\qbar{\overline q}

\def\q5q{\qbar{{\lambda_a}\over 2} i\gamma_5 q}

\def\to{\rightarrow}

\def\a{{\cal A}} 
\def\b{{\cal B}}  
\def\acp{{\cal A}_{\rm CP}}
\def\m{{\cal M}}

\def\he{{\cal H}^{\rm eff}}
\def\l{\lambda}

\def\c{{\cal C}}

\begin{document}

\begin{titlepage}
%
\begin{flushright}
DESY 00-186 \\
December 2000\\
\end{flushright}

\large  
\centerline {\bf Photonic and Leptonic Rare B Decays}
\normalsize

\vskip 2.0cm
\centerline {Ahmed Ali~\footnote{ali@x4u2.desy.de}}
\centerline {\it DESY, Hamburg}
\vskip 4.0cm

\centerline {\bf Abstract}
\vskip 1.0cm
Some selected topics involving photonic and leptonic rare $B$ decays are
reviewed. The interest in their measurement for the
CKM phenomenology is underlined. They are also potentially
interesting in searching for physics beyond the standard model. This is
illustrated on the examples of the decays $B \to (K,K^*) \ell^+ \ell^-$
and $B \to (\rho, \omega) \gamma$ by contrasting their anticipated 
phenomenological profiles in the standard model and some variants of
supersymmetric models.

\vspace*{1.0cm}
\begin{center}
{(\it
Invited Talk; To be published in the Proceedings of the
7th \\International Conference on B Physics at Hadron Machines,\\ Beauty 2000,
Sea of Galilee, Israel, September 13 - 19, 2000.)}
\end{center}
\vfill
\end{titlepage}

\newpage
\section{Introduction}
Rare $B$ Decays (induced in the quark language by transitions such as 
$b \to s \gamma$, $b \to d \gamma$,
$b \to s \ell^+ \ell^-$, $b \to d \ell^+ \ell^- $,...) and
particle-antiparticle mixings ($B^0$ - $\overline{B^0}$, $B_s^0$ - 
$\overline{B_s^0}$,...)  represent flavour-changing-neutral-current
(FCNC) processes. In standard model (SM), FCNC processes
are not allowed in the Born approximation and are induced 
by loops which impart them a sensitivity to higher scales. These
transitions are either dominated by the top quark in the SM, or else the
contribution of the lighter ($u,d,s,c$)-quarks can be estimated using
QCD and the knowledge of the entries in the first two rows
of the CKM (Cabibbo-Kobayashi-Maskawa) matrix \cite{CKM}. In either
case, FCNC $B$ decays can be used to determine the CKM matrix
elements involving the top quark, $V_{td}, ~V_{ts}$, and $V_{tb}$.
Since direct determination of only $\vert V_{tb} \vert$ in the decay
$t \to Wb$ is presently available \cite{cdfvtb}, and there is no credible
method to measure the other two directly, FCNC $B$
(and $K$)-decays are the only feasible alternatives in
quantifying our knowledge about $V_{td}$ and $V_{ts}$.

With the advent of the $B$-factories, the era of 
precision $B$-physics is already upon us. 
With definite plans to accumulate $O(500)$ fb$^{-1}$ luminosity at
the BABAR detector within the next five years (and a similar projection
for BELLE), a large number of rare decays and CP-violating asymmetries 
in partial decay rates of the $B^\pm$ or $B^0/\overline{B^0}$  
mesons will be measured. Hadron
machines will push this frontier even beyond, in particular in the $B_s$ 
and $\Lambda_b$ sectors, as discussed at this meeting
\cite{hadrontalks} and elsewhere \cite{lhcworks}. In anticipation of
this, a lot of theoretical effort has gone in consolidating the
$B$-physics profile in the SM and in suggesting search strategies for
physics beyond the SM.
The prime candidate in new physics searches is supersymmetry. 
While not expected to provide a ''smoking gun'' proof of supersymmetry,
for which high energy colliders are arguably more suited, yet 
precise measurements in low energy processes 
may also reveal effects anticipated in supersymmetric theories. Typical
of these are incremental contributions to the mass differences $\Delta
M_d$ and
$\Delta M_s$ in the $B$ - $\overline{B}$ system, the effective
FCNC vertices $bs\gamma$, $bsg$, $bs \ell^+ \ell^-$, $b s \nu
\bar{\nu}$, inducing the $b \to
s$ transitions (likewise, in $ b\to d$ transitions), and CP-violating
phases.
We review here some selected topics in the photonic and
leptonic rare $B$-meson decays from the point of view of their impact on
the CKM phenomenology and/or their potential for the discovery of 
induced supersymmetric effects.

\section{Radiative Decays $B \to (X_s,X_d) + \gamma$ and the
CKM Phenomenology}
\subsection{${\cal B}(B \to X_s \gamma)$ and determination of $\vert
V_{ts}\vert$}
 We start with the inclusive radiative decays $B \to X_s \gamma$. The
lowest order (one-loop) contribution to the decay $b \to s +\gamma$ is
described by the amplitude
\beq
{\cal M}(b \to s + \gamma) = \frac{G_F}{\sqrt{2}}\frac{e}{2\pi^2}
\sum_{i=u,c,t} \lambda_i F_2(x_i) q^\mu \epsilon^\mu \bar{s} \sigma_{\mu \nu}
\left(m_bR + m_s L \right) b ~,
\eeq
where $ L=(1-\gamma_5)/2$, $R=(1+\gamma_5)/2$, and  $x_i =m_i^2/m_W^2$,
with $m_i$ and $m_W$ being the quark and $W^\pm$-boson masses,
respectively; $G_F$ is the Fermi coupling constant and $e$ is the
electromagnetic coupling, with $e^2/4\pi=\alpha_{em}\simeq 1/137.$
The quantity $F_2(x_i)$ is the Inami-Lim Function \cite{Inami-Lim}
\beq
 F_2(x)=\frac{x}{24(x-1)^4} \left[ 6x(3x-2)\ln x - (x-1)(8x^2 + 5x -7) 
\right] ~,
\eeq
and $\lambda_i$ are the CKM factors, 
$ \lambda_i\equiv V_{ib}V^*_{is}$. Since $\lambda_u/\lambda_c \ll 1$, CKM
Unitarity implies $\lambda_c \simeq -\lambda_t$, yielding the
GIM-amplitude \cite{GIM}: 
\beqa
 {\cal M}(b \to s + \gamma) &=&
 \frac{G_F}{\sqrt{2}}\frac{e}{2\pi^2}\lambda_t \nonumber\\
 && \times \left(F_2(x_t)-F_2(x_c)\right) q^\mu \epsilon^\mu \bar{s} 
\sigma_{\mu \nu} \left(m_bR + m_s L \right) b ~.
\label{GIMmech} 
\eeqa
With $m_t$ already measured \cite{pdg00}, this amplitude depends on
$\lambda_t$.

To quantify this, one has to incorporate QCD corrections to the decay 
rate. A truly cooperative effort by several
theoretical groups has led
to a determination of the branching ratio ${\cal B} ( B \ra  X_{s} \g)$
up to and including the next-to-leading order (NLO) accuracy \cite{greub}. 
The result is the following NLO (in ${\cal O}(\alpha_s)$) and power
(in $1/m_b^2$ and $1/m_c^2$) corrected branching ratio in the SM
\cite{greub,Misiakhep}:
\beq 
 {\cal B} ( B \ra  X_{s} \g) = [(3.29 \pm 0.33)\times 10^{-4}]
\cdot [\vert V_{ts}^*V_{tb}/V_{cb}\vert/0.976]^2~.
\label{bsgamsmbr} 
\eeq
Explicit dependence on the CKM factor $\vert
V_{ts}^*V_{tb}/V_{cb}\vert$  
and the default value following from the (indirect) unitarity fit
\cite{pdg00} of this quantity are displayed. The present world average for
the branching ratio, based on the CLEO \cite{cleobsg}, ALEPH
\cite{alephbsg} and the more recent BELLE \cite{bellebsg} measurements, is
${\cal B} ( B \ra X_{s} \g) =[(3.21 \pm
0.39)\times 10^{-4}$. This yields the CKM ratio:
\beq
\vert\frac{V_{ts}^* V_{tb}}{V_{cb}}\vert = 0.964 ~\pm ~0.075,
\label{vtsbsg}
\eeq
to be compared with the unitarity fits of
this quantity: $\vert\frac{V_{ts}^* V_{tb}}{V_{cb}}\vert = 0.976
~\pm ~0.010$. 
 Using the present measurements of the matrix element
$\vert V_{cb}\vert = 0.04 ~\pm ~0.002$ \cite{pdg00} and assuming 
$\vert V_{tb}\vert \simeq 1.0$,  yields
$\vert V_{ts}\vert = 0.038 ~\pm ~0.003$. Using, instead, the value
$\vert V_{tb}\vert=0.97^{+0.16}_{-0.12}$, measured by the CDF
collaboration \cite{cdfvtb}, gives $\vert V_{ts} \vert =0.04 \pm 0.007$,
which corresponds to a measurement with an error of $\pm 18\%$. This
error will be greatly reduced in future. 

The determination of $\vert V_{ts} \vert$ from the $B \to X_s \gamma$
is occasionally questioned \cite{Misiakhep}.
The issue raised is the following: The dominance of the top-quark
contribution, manifest in the lowest order
GIM-amplitude in Eq.~(\ref{GIMmech}), is no longer present after
including the QCD corrections in the decay rate. This deserves a closer
look. Internal book keeping of the decay rate
for $B \to X_S \gamma$ shows that 
the QCD corrections do not significantly change the intermediate $u\bar{u}$
contribution which remains small and at the level of $2\%$ and 
can be neglected. However, the
QCD-corrected amplitude ${\cal M}(b \to s \gamma)$ receives a very large
contribution from the matrix element of the four-quark operator
$O_2=(\bar{s} \Gamma_i c)(\bar{c} \Gamma_i b)$, involving the intermediate
$c\bar{c}$ loop, proportional to
$\alpha_s(m_b) C_2(m_b) \lambda_c$, where $\lambda_c=V_{cs}^*V_{cb}$. The
CKM dependence of the $c\bar{c}$ contribution to the branching ratio is
hence proportional to $(\lambda_c^2/V_{cb})^2=V_{cs}^2$.
Fortunately, this CKM matrix element is known very precisely thanks to LEP
$\vert V_{cs} \vert=0.989 \pm 0.016$ \cite{LEPVCS}.  Putting in the
other relevant dynamical quantities, the $c\bar{c}$ contribution by itself
gives a branching ratio which is twice the present experimental value of
${\cal B}(B \to X_s \gamma)$ \cite{greub,Misiakhep}. {\it Were it not for
the
top quark, SM would have been ruled out by data on ${\cal B}(B \to X_s
\gamma)$!} Hence, the top qaurk contribution proportional to $\lambda_t$
is absolutely crucial in bringing the SM prediction for this branching
ratio in accord with data. This evidently constrains the CKM ratio $\vert
V_{ts}^*V_{tb}/V_{cb}\vert$ whose present estimate is given above.

\subsection{${\cal B}(B \to X_d \gamma)$ and $a_{CP}(B \to X_d \gamma)$ in
the SM}
 Measuring the CKM-suppressed inclusive radiative decay  $B \to X_d +
\gamma$ will be an experimental {\it Tour de Force}! 
The CKM-allowed ($B \to X_s \gamma$) and the CKM-suppressed ($B \to X_d
\gamma$) decays are anticipated to have rather similar
photon energy spectra \cite{AG92}. However, the
excellent $K/\pi$-separation at the B-factory experiments (and at
CLEO) could be used in
effectively separating the two radiative branches. With an estimated
branching ratio ${\cal B}(B \to X_d + \gamma) = {\cal O}(10^{-5})$
\cite{aag98}, one would need $O(10^8)$ $B\bar{B}$ events, which are
well within reach of the $B$-factories
in the next several years. The experimental effort will yield high
dividends, as one also expects direct CP violation in the decay rates for
$B \to X_d \gamma$ and its charge conjugate $\bar{B} \to X_d \gamma$ which
is measurably large in the SM, with typical estimates being $O(20\%)$
\cite{aag98}. 

 The effective Hamiltonian for the $B \to X_d \gamma$ decays can be
written in the form
\cite{aag98}
\beq
{\cal H}_{eff}(b \to d)=
 - \frac{4 G_{F}}{\sqrt{2}} \, \xi_{t} \, \sum_{j=1}^{8}
C_{j}(\mu) \, \hat{O}_{j}(\mu).
\eeq
The CKM factors are given by $\xi_{j} \equiv V_{jb} \,
V_{jd}^{*},~~j=u,c,t$. 
The Wilson coefficients $C_j(\mu)$ are the same as in the decays $B \to
X_s \gamma$,
but the four-quark operators have now an implicit CKM-parametric
dependence
\beqa
&&\hat{O}_{1} =
 -\frac{\xi_c}{\xi_t}(\bar{c}_{L \beta} \go{\mu} b_{L \alpha}) 
(\bar{d}_{L \alpha} \gu{\mu} c_{L \beta})
 -\frac{\xi_u}{\xi_t}(\bar{u}_{L \beta} \go{\mu} b_{L \alpha})
(\bar{d}_{L \alpha} \gu{\mu} u_{L \beta}) \nonumber \\
&& \hat{O}_{2} =
-\frac{\xi_c}{\xi_t}(\bar{c}_{L \alpha} \go{\mu} b_{L \alpha})  
(\bar{d}_{L \beta} \gu{\mu} c_{L \beta})
 -\frac{\xi_u}{\xi_t}(\bar{u}_{L \alpha} \go{\mu}
b_{L \alpha}) (\bar{d}_{L \beta} \gu{\mu} u_{L \beta}).
\eeqa
 Using the
Wolfenstein parametrization \cite{Wolfenstein}, one has:
\beq    
 \xi_u = A \, \lambda^3 \, (\bar{\rho} - i \bar{\eta}),
~~\xi_c = - A \, \lambda^3 ,
~~\xi_t=-\xi_u - \xi_c ~.
\label{xij}
\eeq
Here, $A \simeq 0.84, ~\lambda=0.22$, but $\bar{\rho}$ and the phase
$\bar{\eta}$ are
only poorly determined, typically lying in the correlated range
$0.05 \leq \bar{\rho} \leq 0.40$ and $0.25 \leq \bar{\eta} \leq 0.50$ (at
95\% C.L.) \cite{al99}. The quantities $\bar{\rho}$ and $\bar{\eta}$ are
the $O(\lambda^2)$
corrected Wolefenstein parameters $\bar{\rho}= \rho(1-\lambda^2/2)$ and
$\bar{\eta}=\eta(1-\lambda^2/2)$ \cite{Burasrhoeta}.
Thus, all three CKM factors in Eq.~(\ref{xij}) are comparable and of order
$\lambda^3$.
This yields the following expression for the branching ratio in the SM:
\beqa           
\label{brxd}
\BBGAMAXD&=&\lambda^2[(1-\bar{\rho})^2 +\bar{\eta}^2] D_{t} \nonumber\\
&\times &\left[1+ \frac{\bar{\rho}^2 + \bar{\eta}^2}{(1-\bar{\rho})^2
+\bar{\eta}^2}\hat{D}_{u} + \frac{\bar{\rho}(1-\bar{\rho}) -\bar{\eta}^2}
{(1-\bar{\rho})^2 +\bar{\eta}^2}\hat{D}_{r}
-\frac{\bar{\eta}}{(1-\bar{\rho})^2
+\bar{\eta}^2} \hat{D}_{i} \right] \, , \nonumber\\
\eeqa
where, typically,  $D_t=3.6 \times 10^{-4}$, $\hat{D}_{u}=0.078$, 
$\hat{D}_{r}=-0.136$, and $\hat{D}_{i}=0.27$ \cite{aag98}.
Using
the range of the  parameters $(\rho,\eta)$ given above yields 
\begin{equation}
 6.0 \times 10^{-6} \leq \BBGAMAXD \leq 2.0 \times 10^{-5},
\label{brbdgama}
\end{equation}

The direct CP-asymmetry in the decay rates defined as
\begin{equation}
\label{asymdef}
a_{CP}(B \to X_d + \gamma)\equiv
 \frac{\Gamma(B \to X_d + \gamma)- \Gamma(\overline{B} \to \overline{X_d}
+
\gamma)}{\Gamma(B \to X_d + \gamma) + \Gamma(\overline{B} \to
\overline{X_d} +
\gamma)}~,
\end{equation}
has not so far been calculated in the NLL precision.
We recall that, as opposed to the decay rate 
$\Gamma(\BBGAMAXD)$, which receives contributions starting from
the lowest order, i.e., terms of the form
$(\alpha^n_s(m_b) \ln^n(m_W/m_b))$, the CP-odd
numerator in Eq.~(\ref{asymdef}) is suppressed by an extra
factor $\alpha_s$, i.e., it starts with terms of the form
$\alpha_s(m_b) \,(\alpha_s^n \log^n(m_W/m_b))$. This
results in a moderate scale dependence of $a_{CP}$.

Using the LL expression for the denominator in eq. 
(\ref{asymdef}), the CP rate asymmetry can be written as \cite{aag98}
\begin{equation}
\label{acpdll}
a_{CP}(\BGAMAXD)=-\frac{Im(\xi_t^*\xi_u)D_i}
{|\xi_t|^2 \, D_t^{(0)}}
=\frac{D_{i} \bar{\eta}}{D_{t}^{(0)} [(1-\bar{\rho})^2 - \bar{\eta}^2]} \,
,
\end{equation}
where $D_t^{(0)}$ stands for the LL part of $D_t$, and its typical
value is $D_t^{(0)} \simeq 2.7 \times 10^{-4}$. Varying the CKM
parameters in the range given above yields:
\beq
 0.10 \leq a_{CP}(B \to X_d + \gamma) \leq 0.35~.
\eeq
Hence, a measurement of ${\cal B}(B \to X_d 
+\gamma)$ and $a_{CP}(B \to X_d + \gamma)$ will have a
big impact on the CKM Phenomenology.

The corresponding CP-asymmetry in the CKM-allowed
decay $B \to X_s \gamma$ in the SM is small, $a_{CP}(\BGAMAXS) \propto
\lambda^2 \eta$ yielding $a_{CP}(\BGAMAXS)\leq 1\%$, reflecting the fact
that the complex phases in the matrix elements $V_{ts}$ and
$V_{tb}$ are absent in leading order. Hence, a measurement of
the CP-asymmetry $a_{CP}(\BGAMAXS)$ will be a sure signal of new physics
\cite{KN98}. Present bounds from CLEO imply that $a_{CP}(\BGAMAXS)$ lies
between $-0.27$ and $+0.10$ at 90\% C.L. \cite{cleobsg}, which is
consistent with being zero and with the SM. It should be stressed that
in the experimental analysis of the CP-asymmetries in inclusive radiative
decays $B \to X \gamma$, it is imperative to separate the hadrons $X$ 
into $X_s$ (emerging from $b \to s$ transitions) and $X_d$ (emerging from
$b \to d$ transitions). Else, as pointed out by Soares \cite{Soares},
the sum of the rate difference $\Delta \Gamma(B \to X_d +\gamma) +
\Delta \Gamma(B \to X_s +\gamma)$ vanishes in the SM in the limit
$m_d=m_s=0$.   

\section{Exclusive Decays $B \to V + \gamma$}
Inclusive decay $B \to X_d \gamma$ is theoretically more robust but
experimentally very challenging to measure. The exclusive decays
$B \to V \gamma$, with $V=\rho^0, \rho^\pm$ and $\omega$, should be 
relatively
easier to measure, given the large statistics at the B factories. In
anticipation of this, considerable effort has gone in studying the
theoretical profile of the exclusive radiative $B$-decays
\cite{brhogam}. We
shall first review some of these estimates in terms of the predicted
branching
ratios for $B \to V \gamma (V=K^*,\rho,\omega)$, which are being used to
constrain the CKM ratio $\vert V_{td}/V_{ts}\vert$ from the current
experiments.

Using the effective Hamiltonian
\beq
{\cal H}(b \to d \gamma)
 =
\frac{G_F}{\sqrt{2}}\left[\xi_{u}(\c_1(\mu) {\cal
O}_1(\mu) + \c_2(\mu) {\cal O}_2(\mu)) -\xi_{t} \c_7^{eff}(\mu)
{\cal O}_7(\mu) +...\right]~,
\eeq
where the Four-quark operators are now defined as:
\beq
{\cal O}_1 = (\bar{d}_\alpha \Gamma^\mu u_\beta)(\bar{u}_\beta
\Gamma_\mu b_\alpha),~~~~ 
{\cal O}_2 = (\bar{d}_\alpha \Gamma^\mu u_\alpha)(\bar{u}_\beta 
\Gamma_\mu b_\beta) ~, 
\eeq 
with $\Gamma_\mu=\gamma_\mu(1-\gamma_5)$,
and ${\cal O}_7 = \frac{e m_b}{8 \pi^2} \bar{d} \sigma^{\mu
\nu}(1-\gamma_5) F_{\mu \nu} b$,
the crucial step is to calculate the form factors 
in $B \to V$ transitions involving these
operators. Defining the form factor in $B \to V + \gamma$ from $O_7$
($\bar{\psi}=\bar{d},~\bar{s}$)
\beq
 \langle V,\lambda |\frac{1}{2} \bar \psi \sigma_{\mu\nu} q^\nu b
 |B\rangle  =
     i \epsilon_{\mu\nu\rho\sigma} e^{(\lambda)}_\nu p^\rho_B p^\sigma_V
F_S^{B\rightarrow V}(0)~,
\eeq
the form factor $B_S^{B \to K^{*}}(0)$ has been measured in the
decays $B \to K^* \gamma$. There are several estimates of this quantity  
based on the light cone QCD-sum rules
\cite{LCQCD-SR,ABHH99} and lattice-QCD \cite{UKQCD}, yielding
branching ratios in agreement
with the CLEO \cite{cleobkstar}, BELLE \cite{bellebsg} and BABAR
\cite{babarbkstar} data.
 
Assuming only the $O_7$ contribution (SD-Dominance) leads to simple
relations
among the decay rates for $B \to \rho \gamma$ and $B \to K^* \gamma$.
For example, one has
\beq
R(\rho/K^*) \equiv \frac{\Gamma (B \to \rho + \gamma)}
     {\Gamma (B \to K^{*} + \gamma)}
  \simeq \kappa_{u,d}\left[\frac{\Vtdabs}{\Vtsabs}\right]^2 ~, 
\eeq
where  $\kappa_{i} \equiv [F_S(B_i \to \rho \gamma)/F_S(B_i \to K^*
\gamma)]^2$ is the ratio of the form factors entering in the SD-part
of the amplitudes. (In the limit of $SU(3)$) symmetry $ \kappa_i
=1$.) Isospin symmetry implies ${\cal B}(B^+ \to \rho^+
\gamma)= 2{\cal B}(B^0 \to \rho^0 \gamma)$ and ${\cal B}(B^+ \to K^{*+}
\gamma)= {\cal B}(B^0 \to K^{0*} \gamma)$. In addition using
SU(3) symmetry, one has ${\cal B}(B^0 \to \omega
\gamma)= {\cal B}(B^0 \to \rho^0 \gamma)$. These relations are frequently
used in the current experimental analyses.   
Present upper limits on $R(\rho/K^*)$ are: $R(\rho/K^*) < 0.32$
\cite{cleobkstar} and $R(\rho/K^*)<0.28$ \cite{bellebsg},  yielding
$\vert V_{td}/V_{ts}\vert<0.75$ and $\vert V_{td}/V_{ts}\vert <0.70$,
respectively, at 90\% C.L.,
using $\kappa_i=0.58$ \cite{abs94}. This is not yet competitive with the 
bound from the ratio on the mass differences $\Delta M_d/\Delta M_s$,
which currently yields
$\vert V_{td}/V_{ts}\vert <0.24$ at 95\% C.L. \cite{al99}.
Experiments at $B$ factories will be able to reach the sensitivity of
the SM, $\vert V_{td}/V_{ts}\vert \simeq 0.2$. 

Next, we discuss the isospin-violating ratio $\Delta$, defined as follows:
\beq
\Delta^{-0} \equiv
  {\Gamma(B^- \to \rho^- \gamma) \over 2 \Gamma(\overline{B^0} \to \rho^0
\gamma)} -1~, ~~\Delta^{+0} \equiv
  {\Gamma(B^+ \to \rho^+ \gamma) \over 2 \Gamma(B^0 \to \rho^0
\gamma)} -1 ~.
\eeq
The charge-conjugate averaged ratio
\beq
\Delta =\frac{1}{2} [\Delta^{-0} + \Delta^{+0}] ~,
\eeq
and the direct CP-asymmetry
\beq
        \acp (B^\pm \to \rho^\pm \gamma) \equiv
        \frac{\b(B^- \rightarrow \rho^- \gamma)
          - \b(B^+ \rightarrow \rho^+ \gamma)}{
          \b(B^- \rightarrow \rho^- \gamma)
          + \b(B^+ \rightarrow \rho^+ \gamma)} \; ~,
\eeq
(likewise, $\acp (B^0 \to \rho \gamma)$) are the quantities of
principal interest in this talk.

 The isospin-breaking effects in the ratio $\Delta^{+0}$ and
$\Delta^{-0}$ due to the long-distance contributions in the decays
$B^\pm \to \rho^\pm \gamma$ may turn out to be sizable. Hence,
given enough data, one should
analyze the ratios $\Delta^{+0}$ and $\Delta^{-0}$ for anticipated 
deviations from the isospin value of unity. In fact, as argued in
Ref.~\cite{AHL00}, the interference of the LD and SD-amplitudes
in $\Delta$ and  $\acp(B^\pm \to \rho^\pm)$  may 
provide a sensitive probe for physics beyond the SM.
  
  The long-distance
contributions arise from the matrix elements of the 
operators ${\cal O}_1$ and   ${\cal O}_2$. 
There are two form factors in the matrix elements
$\langle V,\lambda | {\cal O}_{1,2} | B \rangle$,
 which we denote as $F_1^{L}(q^2)$ and
$F_2^{L}(q^2)$. QCD sum rule-based estimates put them almost equal
\cite{ab95,KSW95}, $F_{1,L}^{B \to V}(q^2)\simeq
F_{2,L}^{B \to V}(q^2)=F_{L}^{B \to V}(q^2)$.
Taking ino account the dominant LD-contributions
arising from  $W^\pm$-annihilation and $W^\pm$-exchange,
one has
\beqa
{\cal M}(B^- \to \rho^- \gamma) &=& \xi_t
a_{P}^{(-)}(1-\frac{\vert \xi_u\vert}{\vert
\xi_t \vert}
R_{L}^{(-)} e^{i\alpha})~, \nonumber\\
{\cal M}(\overline{B^0} \to \rho^0 \gamma) &=& \xi_t
a_{P}^{(0)}(1-\frac{\vert
\xi_u\vert}{\vert\xi_t\vert}
R_{L}^{(0)} e^{i\alpha})~, 
\eeqa
where $\alpha$ is one of the inner angles of the unitarity
triangle; $a_P^{(-)}$ and $a_{P}^{(0)}$
are SD (Penguin)-amplitudes and $R_{L}^{(\pm)}$ and $R_{L}^{(0)}$ are
model-dependent quantities involving the form factors and other
dynamical variables.
Typical estimates obtained using   
factorization of the matrix elements of ${\cal O}_1$ and ${\cal O}_2$ and
the Light-cone QCD sum rules are \cite{ab95,KSW95}:
\beq
a_P^{(-)} \simeq a_{P}^{(0)}, ~~~R_L^{(-)} \simeq -0.3\pm 0.07,
~~~R_L^{(0)} \simeq 0.03 \pm 0.01~. 
\eeq
The quantity $R_{L}^{(0)}$ entering in $\overline{B^0} \to \rho^0 \gamma$
decay is suppressed due to the electric charge of the down quark
($e_{\bar{d}}=+1/3)$ in
$\overline{B^0}=b\bar{d}$, as opposed to the up quark charge
($e_{\bar{u}}=-2/3)$
in $B^-=b\bar{u}$, as radiation from the light quarks dominates, and the
color-suppression factor, which phenomenologically has a relative weight
of about 0.25.

Dropping $R_L^{(0)}$ and rewriting $R_L^{(-)} =\epsilon_A {\rm e}^{i
\phi_A}$, where $\phi_A$ is a strong interaction phase, on has in the
lowest order
\beq
   \frac{\b(B^- \rightarrow \rho^- \gamma)}{2
     \b(\overline{B^0} \rightarrow \rho^0 \gamma)} \simeq  
     \left| 1 - \epsilon_A {\rm e}^{i \phi_A}
\frac{\vert\xi_u\vert}{\vert\xi_t\vert} e^{+i\alpha} \right|^2 ~.
\eeq
The isospin ratios can now be expressed as:
\beq
  \Delta^{\pm 0} = 2 \epsilon_A \left[
    \cos \phi_A F_1 \mp \sin \phi_A F_2 +
    \frac{1}{2} \epsilon_A (F_1^2 + F_2^2)
    \right] ~.
\eeq
Here, 
$F_{1,2}$ are  functions of the Wolfenstein parameters
$\bar{\rho}$ and $\bar{\eta}$, with
\beq
F_1= - \left\vert {\l_u^{(d)} \over \l_t^{(d)}}\right\vert \cos\alpha,~~~ 
F_2= - \left\vert {\l_u^{(d)} \over \l_t^{(d)}} \right\vert \sin\alpha, 
~~~(F_1^2 + F_2^2) = \left\vert \l_u^{(d)} / \l_t^{(d)}\right\vert^2 ~.
\eeq
The charge-conjugated ratio $\Delta$ in the leading order is:
\beq
 \Delta_{\rm LO} = 2 \epsilon_A \left[\cos \phi_A F_1 + \frac{1}{2}
\epsilon_A(F_1^2 + F_2^2)\right]
\simeq 2 \epsilon_A \left[F_1 +{1 \over 2}\epsilon_A(F_1^2
+F_2^2) \right]~,
\label{deltalo}
\eeq
the second equality follows from assuming $\phi_A=0$, obtaining in the
factorization approximation.

Recently, the leading-twist non-factorizable contribution
to the weak annihilation amplitude has been computed in perturbation
theory by Grinstein and Pirjol, using heavy quark expansion techniques
\cite{GP00}.
In the chiral limit $(m_u,m_d \to 0)$, these authors show  
that the one-loop non-factorizable corrections in weak
annihilation amplitude vanish in the leading twist
and one recovers the factorization result. Interestingly,
the weak annihilation contribution, which is the dominant
long-distance amplitude, can
be determined  model independently from the  radiative decays $B^\pm \to
\gamma \ell^\pm \nu_\ell$ \cite{KPY00}.
To generate non-zero ${\cal A}_{\rm CP}$, it is necessary to compute
the NLO corrections \cite{Soares,GSW}, which generates $\phi$
perturbatively in the penguin amplitude. Of course, 
$\Delta_{\rm LO}$ is also renormalized by perturbative QCD corrections.

\subsection{NLO Effects in $\Delta$ and ${\cal A}_{CP}(B^\pm \to
\rho^\pm \gamma)$}

\begin{figure}[t!]
      \centering
     \begin{minipage}[c]{0.5\textwidth}
       \centering
       \includegraphics[scale=0.5]{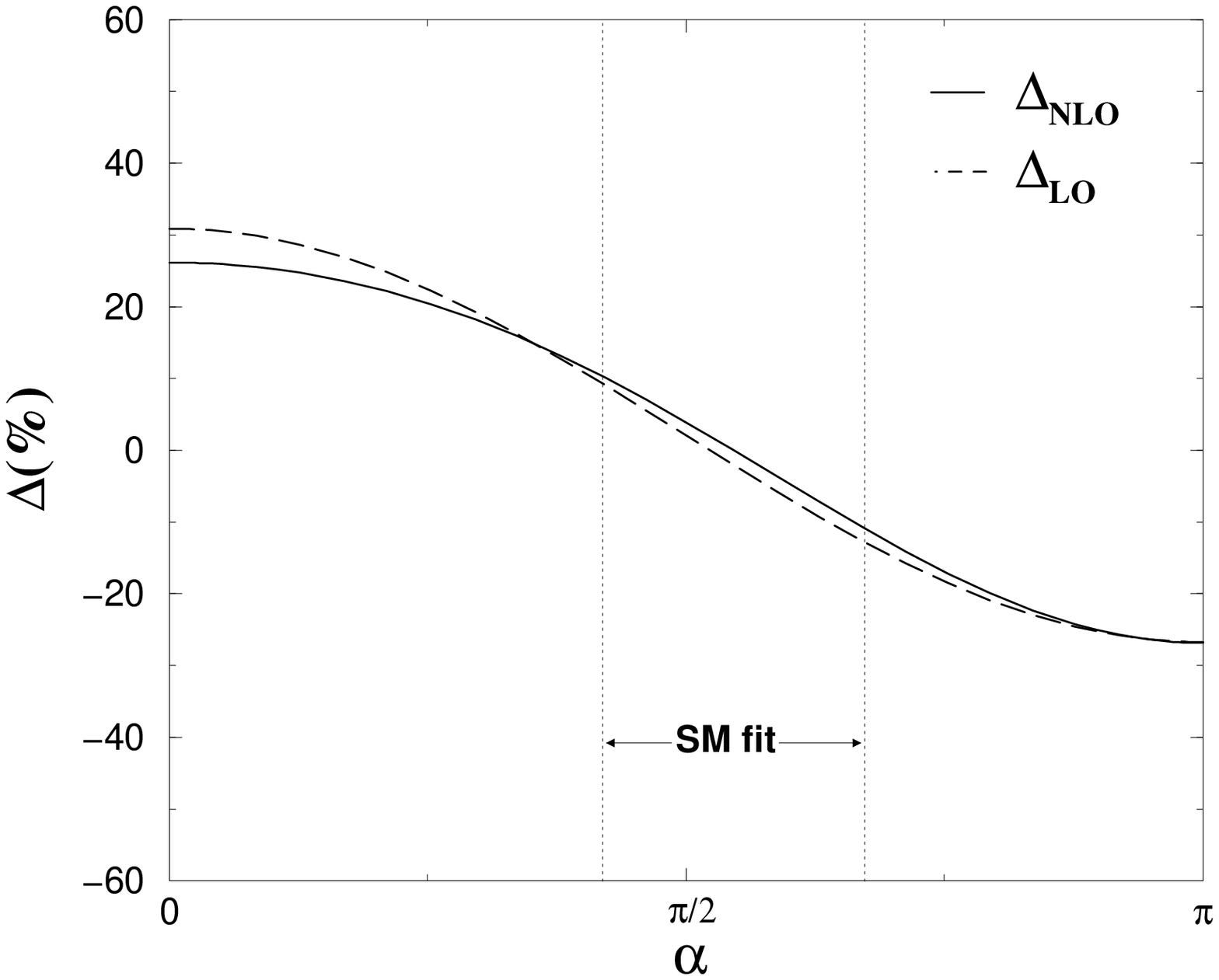}\\
     \end{minipage}
\caption[]{$\Delta_{\rm LO}$ and $\Delta_{\rm NLO}$
      in SM with $\epsilon_A=-0.3$ and $\vert V_{ub} / V_{td} 
      \vert=0.48$.
(From Ref.~\protect\cite{AHL00}.)}
\label{fig:delta}
\end{figure}

Ignoring contributions which are formally power ($1/m_b$)-suppressed, the
NLO corrections to the decay rate
$\Gamma(B \to \rho \gamma)$ can be retrieved from the corresponding
corrections to the inclusive radiative decays $\Gamma(B \to X_s \gamma)$ 
and $\Gamma(B \to X_d \gamma)$. The
isospin-violating ratio $\Delta_{\rm NLO}$ in the NLO approximation and
the CP-violating decay rate asymmetry $\acp(B^\pm \to \rho^\pm \gamma)$
have been calculated to $O(\alpha_s)$ in Ref.~\cite{AHL00}:
 The isospin-violating ratio $\Delta_{\rm NLO}$
and $\Delta_{\rm LO}$ are shown as functions of the angle $\alpha$ in
Fig.~\ref{fig:delta}. As can be seen in this figure, the NLO corrections
to $\Delta_{LO}$ are small 
for all values of $\alpha$. The anticipated range of $\alpha$
from the CKM-unitarity fits with 95\% C.L.\cite{al99} is also indicated.
Note, $\Delta_{\rm LO}$ (but numerically also
$\Delta_{\rm NLO}$) is essentially proportional to $F_1$, while
 $\acp(B^\pm \to \rho^\pm \gamma)$ is
proportional to $F_2$. Thus, for the central value of the
CKM-unitarity fits, yielding $\alpha \simeq \pi/2$, ~$\Delta_{\rm NLO}$ is
very small, implying that isospin symmetry in $\Delta$ holds to a
very high precision, while for this value of $\alpha$,
$\acp(B^\pm \to \rho^\pm \gamma)$ takes its maximum
value. Conversely, if the value of $\alpha$ is sufficiently different from
$\pi/2$, the CP asymmetry $\acp(B^\pm \to \rho^\pm \gamma)$ decreases
while $\Delta_{\rm NLO}$ increases, reaching $\Delta_{\rm NLO} \simeq \pm
0.2$
for extreme allowed values of $\alpha$. Thus, measurements of $\Delta$
and $\acp(B^\pm \to \rho^\pm \gamma)$ determine the angle $\alpha$ in the
SM \cite{AHL00}, providing complementary information on this angle.

\begin{figure}
      \centering
     \begin{minipage}[c]{0.4\textwidth}
       \centering
       \includegraphics[scale=0.4]{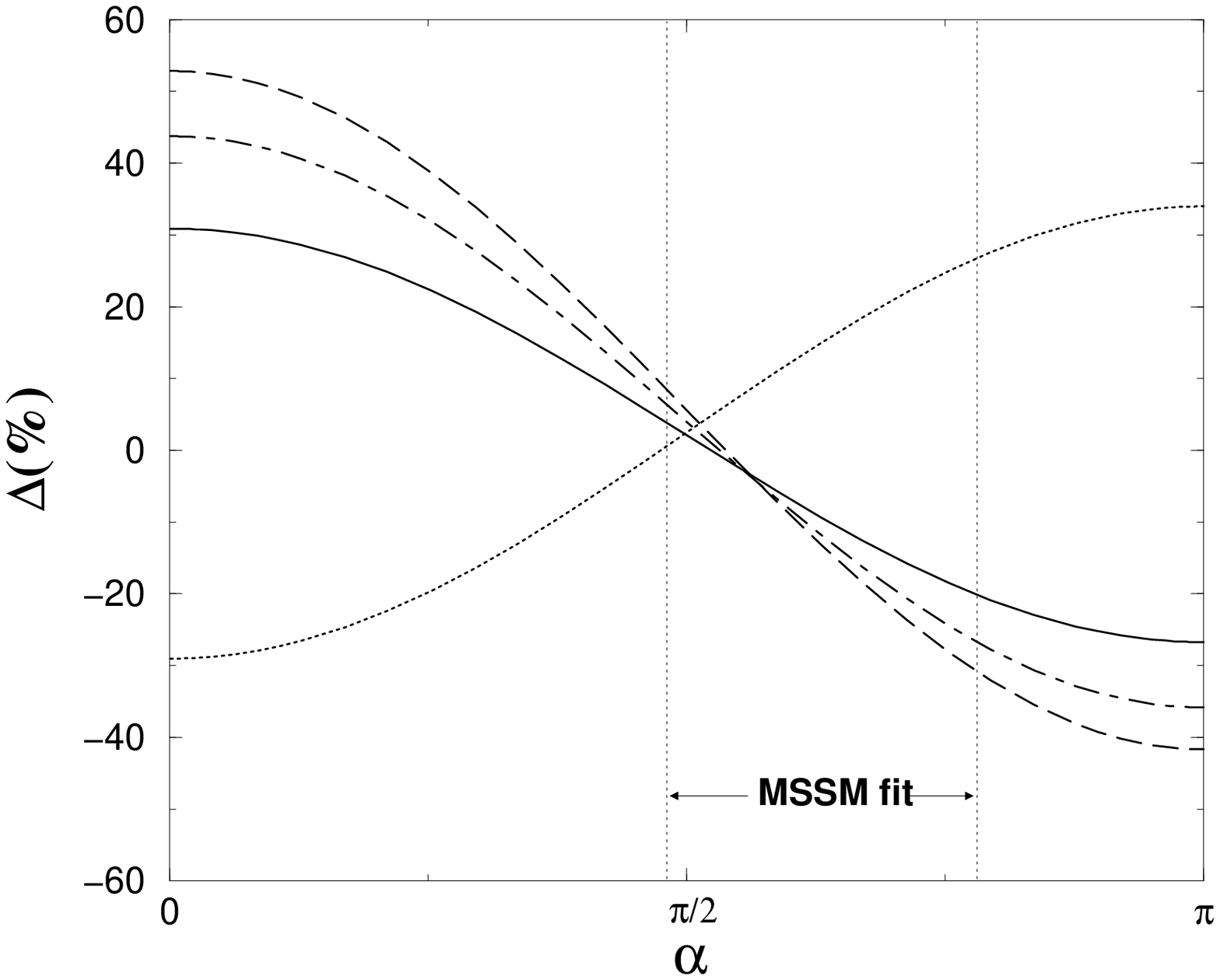}
     \end{minipage}
     \hspace*{1cm}
     \begin{minipage}[c]{0.4\textwidth}
       \centering
       \includegraphics[scale=0.4]{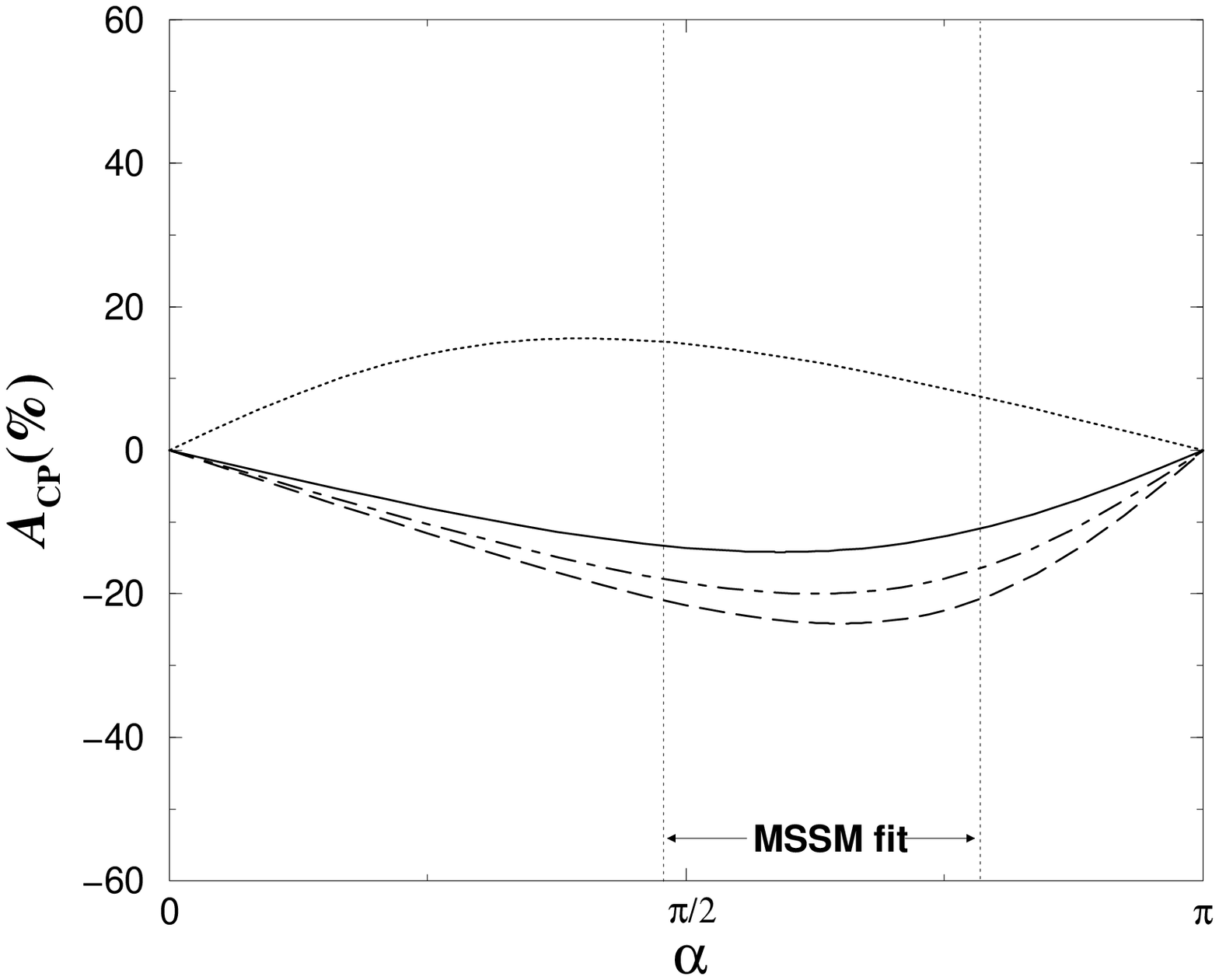}
     \end{minipage}
\caption[]{$\Delta_{\rm LO}$ (left) and
      $\acp(B^\pm \to \rho^\pm \gamma)$ (right) with $\epsilon_A=-0.3$
      in the SM (solid
      line), and in the MSSM with
      $\c_7^{(0)eff}/\c_7^{(0)eff\rm (SM)} = 0.95$ (dot-dashed
      line), $ \c_7^{(0)eff}/\c_7^{(0)eff\rm (SM)} =
      0.8$ (dashed line) and
      $\c_7^{(0)eff}/\c_7^{(0)eff\rm (SM)} = -1.2$ (dotted
      line). The SM and MSSM curves
      correspond respectively to $\vert V_{ub} / V_{td} \vert=0.48$
      and $\vert V_{ub} / V_{td} \vert=0.63$. 
      (From Ref.~\protect\cite{AHL00}.)}
\label{deltasusy}
\end{figure}

As a possible candidate for new physics, we discuss supersymmetric effects
on $\Delta$ and ${\cal A}_{\rm CP}$ which enter through the modification of
the Wilson coefficient $C_7^{(0)eff}(m_b)$
and some $O(\alpha_s)$ functions (called $A_R^{(1)t}$ and $A_I^{(1)t}$
in Ref.~\cite{AHL00}). In
addition, $F_1$ and $F_2$ are also in general modified.
 We shall restrict ourselves to the minimal supersymmetric model (MSSM)
case, in which one expects
additional contributions to $\Delta M_d$, $\Delta M_s$, and $\epsilon_K$,
but no new phases, leading to a shift in the apex of the CKM-unitarity
triangle.
The branching ratio ${\cal B}(B \to X_s \gamma)$ constrains the real and
imaginary parts of $C_7^{eff}$. Including upper bounds from the electric
dipole moment of the neutron leads to ${\rm Im} (C_7/C_7^{\rm SM}) \ll 1$
\cite{goto99,bksusy}.
Depending on the supersymmetric parameters, one has typically three
different situations:
(i) Small $\tan \beta$-solution
(say, $\tan \beta =3$): ${\rm Re}(C_7^{eff}/C_7^{eff,{\rm SM}})=1 \pm 15\%
$;
(ii)Medium $\tan \beta$-solution (say, $\tan \beta =10$):  
${\rm Re}(C_7^{eff}/C_7^{eff,{\rm SM}})=0.7$--$1.2$;
(iii) Large $\tan \beta$-solution (say, $\tan \beta =30$). In the last 
case, two possible branches 
are: either ${\rm Re}(C_7^{eff}/C_7^{eff,{\rm SM}})=0.7$--$1.2$, or
${\rm Re}(C_7^{eff}/C_7^{eff,{\rm SM}})=(-0.8)$--$(-1.5)$. It is this
second
possibility involving a flip of the sign of $C_7^{eff}$, compared to the
SM, which is of particular
interest in the context of the isospin-violating ratio $\Delta_{\rm NLO}$
and the CP asymmetry $\acp(B^\pm \to \rho^\pm \gamma)$. The interference
of the SD-
and LD-contribution in the amplitude, impacting on $\Delta$ and
$\acp(B^\pm \to \rho^\pm \gamma)$, is indeed sensitive to the sign of
$C_7^{eff}$. Possible supersymmetric
effects on $\Delta$ and $\acp(B^\pm \to \rho^\pm \gamma)$ are shown in
Figs.~\ref{deltasusy},
where the assumed values of ${\rm Re}(C_7^{eff}/C_7^{eff,{\rm SM}})$ are
also indicated. 
%
%
\section{\bf The decays $B \to (K,K^*) \ell^+ \ell^-$ in the SM}
\subsection{The effective Hamiltonian approach}
The decays $B \to (X_s,X_d) \ell^+ \ell^-$, as well as their
exclusive counterparts such as the decays $B \to
(K,K^*,\pi,\rho) \ell^+ \ell^-$, provide the possibility
of measuring Dalitz-distributions in a number of variables
from which the effective vertices in the 
underlying theory can be extracted \cite{agm95}.
Inclusive decays are theoretically more robust 
than exclusive decays which require additionally the knowledge
of the relevant form factors. However, despite comparatively
larger branching ratios, inclusive decays are more
difficult to measure, in particular, in experiments operating at hadron
machines. Exclusive semileptonic decays are accessible to a wider variety of
experiments \cite{cdfexcl,cleoexcl}. We review here
the anticipated profiles for the decays $B \to (K,K^*) \ell^+ \ell^-$
in the SM and some variants of supersymmetry.

At the quark level, the semileptonic decay 
$b \to s  \ell^+ \ell^-$ can be described in terms of the effective 
Hamiltonian  
\begin{equation}
        \he = -4 \frac{G_F}{\sqrt{2}}  V_{t s}^\ast  V_{tb}  
              \sum_{i=1}^{10} \c_i(\mu)  \o_i(\mu) \; , 
        \label{eq:he}
\end{equation}
where unitarity of the CKM matrix and the numerical hierarchy
$V_{u s}^\ast  V_{ub} \ll V_{t s}^\ast  V_{tb}$ are implicit.
The definitions of the operators and the expressions for the Wilson
coefficients in the SM can be seen in Ref.~\cite{burasmuenz}.
Restricting ourselves to the SM and SUSY, the short-distance contributions
in the decays $B \to X_s \gamma$ and $B \to X_s \ell^+ \ell^-$,
and the exclusive decays of interest to us, are
determined by three coefficients, called  $\cse\equiv C_7-C_5/3 -C_6$,
$\cn$ and $\ct$  
\cite{burasmuenz}.\footnote{In general, more operators are 
present in supersymmetric theories and their possible effects
are discussed in Ref.~\cite{ABHH99}.}

The above Hamiltonian leads to the following free quark decay amplitude
for $b \to s \ell^+ \ell^-$: 
\begin{eqnarray}
        \m(b\to s\ell^+\ell^-) & = & \frac{G_F \alpha}{\sqrt{2}  \pi} \, 
                V_{t s}^\ast V_{tb} \, \left\{
                  \cne  \left[ \bar{s}  \g_\mu  L  b \right] \, 
                          \left[ \lb  \g^\mu  \l \right]
                + \ct  \left[ \bar{s}  \g_\mu  L  b \right] \, 
                         \left[ \lb  \g^\mu  \g_5  \l \right]
                \right. \nonumber \\
        & & \; \; \; \; \; \; \; \; \; \; \; \; \; \; 
            \; \; \; \; \; \; \; \; \; \left. 
                - 2 \mbh  \cse  \left[ \bar{s}  i  \sigma_{\mu \nu}  
                        \frac{\qh^{\nu}}{\sh}  R  b \right] 
                        \left[ \lb  \g^\mu  \l \right]
                \right\} \; .
        \label{eq:m}
\end{eqnarray}
Here, $s = q^2$, $q=p_{+} +p_{-}$, 
$p_{\pm}$ are the four-momenta of 
the leptons, respectively, and the hat on a variable denotes its 
normalized value with respect to the $B$-meson mass, $m_B$, e.g.,
$\sh=s/m_B^2$, $\mbh=m_b/m_B$, and we denote by $m_b \equiv m_b(\mu)$
the $\overline{\rm MS}$ mass evaluated at a scale $\mu$.
Note that $\m(b \to s \ell^+ \ell^-)$, although a free quark decay 
amplitude, contains certain long-distance effects from the matrix elements
of  four-quark operators, $\langle \ell^+ \ell^- s | {\cal O}_i 
 | b \rangle$, $1\leq i \leq 6$, which usually are absorbed into
a redefinition of the short-distance
Wilson-coefficients. Thus, the effective
coefficient of the operator ${\cal O}_9$ is defined as  
\begin{equation}
\cne (\hat{s}) = C_9 + {Y} (\hat{s}) \; ,
\label{eqn:c9eff}
\end{equation}
where $Y(\sh)$ stands for the
above-mentioned matrix elements of the four-quark operators. 
A perturbative calculation yields \cite{burasmuenz}:
\begin{eqnarray} 
        {Y}_{\rm pert} (\sh) & = & g(\mc,\sh) C^{(0)}
         -\frac{1}{2} g(1,\sh)
                \left( 4 \, C_3 + 4 \, C_4 + 3 \,
                C_5 + C_6 \right) \nonumber\\ 
        & - & \frac{1}{2} g(0,\sh) \left( C_3 +   
                3 \, C_4 \right) 
        +      \frac{2}{9} \left( 3 \, C_3 + C_4 +
                3 \, C_5 + C_6 \right) \; , 
\label{eq:y}
\end{eqnarray}
with $C^{(0)} \equiv 3 C_1 + C_2 + 3 C_3 + C_4 + 3 C_5 + C_6$.
The functions $g(\hat{m}, \hat{s})$ are given in Ref.~\cite{burasmuenz}.
Thus, whereas the coefficients $C_7^{eff}$ and $C_{10}$ are constants,
the effective coefficients $C_9^{eff}$ is an $\hat{s}$-dependent function
and has a non-local character. 

 Nonperturbative effects originate in particular from resonance
contributions. It is essentially only the $c\bar{c}$-resonant contribution 
that matters. Ref.~\cite{amm91} suggests to add the 
contributions from $J/\Psi, \Psi^\prime,\dots$ 
to the perturbative result, with the former parametrized in the form of
Breit-Wigner functions with known widths. The function $Y(\hat{s})$ is then
replaced by
\begin{equation}
Y_{\rm amm}(\sh) = Y_{\rm pert}(\sh) + \frac{3 \pi}{\alpha^2} C^{(0)}
         \sum_{V_i = \psi(1s),..., \psi(6s)} \Omega_i
      \frac{\Gamma(V_i \rightarrow \ell^+ \ell^-)\, m_{V_i}}{
      {m_{V_i}}^2 - \sh \, {m_B}^2 - i m_{V_i} \Gamma_{V_i}}~.
\label{eq:amm}
\end{equation}
The phenomenological factors $\Omega_i$ can be fixed from the relation 
\begin{eqnarray}
{\cal{B}}(B\to K^{(\ast)} V_i \to K^{(\ast)} \ell^+ \ell^-)=
{\cal{B}}(B\to K^{(\ast)} V_i) {\cal{B}}(V_i \to\ell^+ \ell^-) \; ,
\end{eqnarray}
where the right-hand side is given by data \cite{pdg00}.

In Ref.~\cite{ks96}, Kr\"uger and Sehgal have argued that 
the perturbative/non-perturbative dichotomy in $Y(\hat{s})$ can be avoided
by 
using the measured cross-section $\sigma(e^+ e^-\to\,$hadrons) 
together with the assumption of the quark-hadron duality for large $\sh$
to reconstruct $Y(\sh)$ from its imaginary part by a dispersion relation.
In principle, this approach has a merit. 
However, the quark-hadron duality argument is not completely
quantitative either, as perturbative contributions in $\sigma(e^+
e^-\to\,$hadrons) and $ {\cal B}(B \to X_s \ell^+ \ell^-)$ are not
identical.
In particular, the (non-local) perturbative part of $Y(\sh)$ has genuine
hard
contributions proportional to $m_b^2$, which can neither be ignored nor
taken care of by the $e^+ e^-$ data. The issue remains as to how much of the
genuine perturbative contribution 
in $ B \to X_s \ell^+ \ell^-$ arising from the $c\bar{c}$-continuum should be
kept and there is at present no unique solution to this problem. Luckily,
the inherent theoretical uncertainty in the two approaches is not
overwhelming and the SD-contribution can be meaningfully extracted from
data.

\subsection{Dilepton invariant mass and the forward-backward asymmetry}

Exclusive decays $B\to (K,K^*) \ell^+ \ell^-$ are
described in terms of matrix elements of the quark operators in
Eq.~(\ref{eq:m}) over meson states, which can be parametrized in terms of form
factors. For $B \to K$ transition, there are three of them, called
$f_{+}(s)$, $f_{0}(s)$, involving the vector current, and
$f_T(s)$, entering in the matrix element of the tensor current.
For the vector meson $K^*$,
there are five form factors in the $V-A$ current, but only four
are independent which we call as $A_1,~A_2,~A_0$ and $V$. The tensor part
involves three form factors, called $T_1$, $T_2$ and $T_3$ \cite{ABHH99},
with $T_1(0)=T_2(0)$. 

 The dilepton mass spectrum for the decays $B \to (K,K^*) \ell^+ \ell^-$
and the forward backward asymmetry (FBA),  defined as \cite{amm91}
\begin{equation}
  \frac{d \a_{\rm FB}}{d \sh} = 
        -\int_0^{\uh(\sh)} d\uh \frac{d^2\gl}{d\uh d\sh}
              + \int_{-\uh(\sh)}^0 d\uh \frac{d^2\gl}{d\uh d\sh} \; ,
  \label{eq:dfba}
\end{equation}
where the variable $\uh$ corresponds to $\theta$, the angle between
the momentum of the $B$ meson and the positively charged lepton
$\ell^+$ in the dilepton center of mass system, are
worked out in a number of papers (see, for
example, Ref.~\cite{ABHH99}).
Assuming CP-symmetry, the forward-backward asymmetry in the
decays of the $B$ and $\bar{B}$ mesons are equal and opposite
\cite{BHI01}.
 
For $B \to K$ transitions, the dilepton invariant mass spectrum  is
simplified
\begin{equation}
\frac{d\Gamma} {d\hat{s}} \sim \vert V_{ts}^*V_{tb} \vert^2 \left(
\vert C_9^{eff} f_{+}(s) + \frac{2 \hat{m}_b}{1+\hat{m}_K}
C_7^{eff}f_T(s)\vert^2 + \vert C_{10}f_{+}(s)\vert^2 \right) ~.
\label{btokdilept}
\end{equation}
Note, there is no dependence
on the form factor $f_{-}$ for $m_\ell =0$. Also, since $\vert
C_7^{eff}\vert \ll \vert  C_9^{eff}\vert, \vert C_{10}\vert$, to a good
approximation $d\Gamma/d\hat{s}
\propto \vert f_{+} \vert^2$, with the effect from the $C_7^{eff} f_T$
term of order $-10\%$. Hence, the $B \to K \ell^+ \ell^-$ decay is
dominated by the
matrix element of the vector current, just like the charged
current induced transition $B \to \pi \ell \nu_\ell$. 
This observation could be put to good use to determine
the CKM ratio $\vert V_{ub}/V_{ts}^*V_{tb} \vert$ \cite{lsw98}.
The FBA vanishes in $B\to K\ell^+\ell^-$ decays. 

  For the $B \to K^* \ell^+ \ell^-$ transition, there is no dependence
on the form factor $A_0$ for $m_\ell=0$. However,  in general, 
the dilepton invariant mass distribution depends on all the three 
effective Wilson coefficients and the form factors discussed
above \cite{ABHH99}.The expression for the FBA for the decay $B\to
K^*\ell^+\ell^-$ is rather simple and instructive
\begin{eqnarray}
  \frac{d \a_{\rm FB}}{d \sh}& =& 
  \frac{G_F^2 \, \alpha^2 \, m_B^5}{2^{8} \pi^5} 
      \left| V_{ts}^\ast  V_{tb} \right|^2 \, \sh \uh(\sh)^2 \nonumber \\
& & \times  \ct 
\left[  {\rm Re}(\cne) V A_1+ \frac{\mbh}{\sh} \cse (V T_2 (1-\mvh)+
A_1 T_1 (1+\mvh)) \right] \; ~. \nonumber\\
  \label{eq:dfbabvllex}
\end{eqnarray}
The position of the zero $\sh_0$ of the FBA is given by
\begin{eqnarray}
{\rm Re}(\cne(\sh_0)) =- \frac{\mbh}{\sh_0} \cse 
\left\{\frac{T_2(\sh_0)}{A_1(\sh_0)} (1-\mvh)+
\frac{T_1(\sh_0)}{V(\sh_0)} (1+\mvh)\right\} \; ,
\label{eq:fbzero}
\end{eqnarray}
which depends on the value of $m_b$, the ratio 
of the effective coefficients $\cse/{\rm Re}(\cne(\sh_0))$, and  
the ratio of the form factors shown above.
It is interesting to observe that using the 
heavy quark symmetry, formulated in the form of a Large Energy Effective 
Theory (LEET) \cite{LEET}, both the ratios of
the form factors appearing in 
Eq.~(\ref{eq:fbzero}) can be shown to have essentially no hadronic
uncertainty, i.e.\,, all dependence on the form factors cancels:
\begin{equation} 
\frac{T_2}{A_1}= \frac{1+\mvh}{1+\mvh^2-\sh} 
\left(1-\frac{\sh}{1-\mvh^2}\right) \; ,
~~~\frac{T_1}{V}  = \frac{1}{1+\mvh} \; .\label{eq:FBA}
\end{equation}
With these relations, one has a particularly simple form for the 
right hand side in Eq.~(\ref{eq:fbzero}) determining $\sh_0$, namely
\begin{equation}
{\rm Re}(\cne(\sh_0)) =- 2 \frac{\mbh}{\sh_0} \cse
\frac{1-\sh_0}{1+m_{K^*}^2 -\sh_0} \; .
\label{eq:fbzeroleet}
\end{equation}
Thus, the zero-point of the FB-asymmetry in $B \to K^* 
\ell^+ \ell^-$ is determined essentially
by the precision of the ratio of the effective coefficients and $m_b$, 
putting it on almost same footing as the corresponding quantity in the
inclusive
decays $B \to X_s \ell^+ \ell^-$, for which the zero-point of the FBA is
given by
the solution of the equation ${\rm Re}(\cne(\sh_0)) = -  
\frac{2}{\sh_0} \cse$.  
The insensitivity of $\sh_0$ to the decay form factors in $B \to K^* 
\ell^+ \ell^-$ has also been pointed out in Ref.~\cite{burdman}
based on a comparative study of a number of form factor models. 
Perturbative stability of the LEET results given in
Eqs.~(\ref{eq:FBA}) has recently been studied by Beneke and
Feldmann
\cite{BF00}, who find that the $O(\alpha_s)$ corrections, modifying the
above relations,  are within $10\%$. Thus, the LEET-based
result in Eq.~(\ref{eq:fbzeroleet}) stands theoretically on quite rigorous
grounds, making the zero-point of the FBA a precision test of the SM. 
Typically, $\sh_0=0.10$ (i.e.\ $s_0=2.9\, \mbox{GeV}^2$) in the SM.

 While none of the
experiments has so far reached the SM-sensitivity
in $B \to (K,K^*) \ell^+ \ell^-$ decays, some do provide
interesting upper limits on the parameter space of models with new
physics. This has been worked out in the context of the SUSY models        
in Ref.~\cite{ABHH99}. The exclusive
decay $B \to K^* \mu^+ \mu^-$ provides at present the most
stringent bounds on the effective coefficients in the SM. The current
upper limit \cite{cdfexcl} and the SM-expectations \cite{ABHH99} are:
\begin{eqnarray}
{\cal B}_{nr}(B \to K^* \mu^+ \mu^-)_{\mbox{SM}} &=& 2.0 \pm 0.5 \times
10^{-6}~, \nonumber\\
{\cal B}_{nr}(B \to K^* \mu^+ \mu^-)_{\mbox{CDF}} & < & 4.0 \times
10^{-6} ~~(\mbox{at 90\% C.L.})~.
\label{eq:bkmumusm}
\end{eqnarray}
\section{The decays $B \to (K,K^*) \ell^+ \ell^-$ in SUSY}
Rare $B$-decays $B \to X_s \gamma$ and $B \to X_s
\ell^+ \ell^-$ have been extensively studied in the context of
supersymmetric theories. Recent developments can be seen in
Refs.~\cite{goto99,bksusy,LMSS99,goto96,MFVbsg,bghw00}.
We do not consider models with broken R-parity and assume that there are no 
new phases from {\it new physics} beyond the SM. This covers an important
part of the supersymmetric parameter space, but not all. 

Possible contributions from new physics (NP) in the  
relevant Wilson coefficients can be taken into account by the (correlated) 
ratios, ($i=7,9,10$):
\begin{equation}
R_i(\mu)\equiv \frac{{\cal{C}}^{NP}_i+{\cal{C}}^{SM}_i}
{{\cal{C}}^{SM}_i}=\frac{{\cal{C}}_i}{{\cal{C}}^{SM}_i} \; ,
\end{equation}
which depend on the renormalization scale (except for $\ct$). The
experimental constraint from $B \to X_s \gamma$ translates into the bound
\begin{equation}  
0.80 < |R_7(\mu=4.8~\mbox{GeV})| < 1.20~,
\label{eq:R7bounds}
\end{equation}
where the coefficients are understood to be calculated in the LLA precision.
Some representative cases are discussed below.  
\subsection{$B \to (K,K^*) \ell^+ \ell^-$ in SUGRA models}
The parameter space of these models may be decomposed into two qualitatively 
different regions, which can be characterized by $\tan\beta$ values.
 For small $\tan\beta$, say $\tan\beta \sim 3$, the sign of 
$\cse$ is the same as in the SM. Here, no spectacular deviations from the 
SM can be expected in the decays $B \to (K,K^*) \ell^+ \ell^-$.
For large $\tan\beta$, the situation is more interesting due to correlations
involving the branching ratio for $B \to X_s \gamma$, the mass of the 
lightest CP-even Higgs boson, $m_h$, and sign$(\mu_{susy})$, appearing in 
the Higgs superpotential. The interesting scenario for
SUSY searches in $B \to (K,K^{*}) \ell^+ \ell^-$ is the one in which 
sign$(\mu_{susy})$ and $m_h$ admit $\cse$ to be positive
\cite{goto99}. In this case one expects 
a constructive interference of the terms depending on $\cse$ and $\cn$ in 
the dilepton invariant mass spectra. For the sake of illustration, we 
use
\begin{equation}
R_7=-1.2,~~R_9=1.03,~~R_{10}=1.0 ~,
\label{eq:sugrar}
\end{equation}
obtained for 
$\tan\beta= 30$ \cite{goto96}, as a representative large-$\tan\beta$
solution.

In Fig.~\ref{fig:BKsusy}, we show the dilepton invariant mass spectrum for
the decay $B \to K \ell^+ \ell^-$. Below the $J/\psi$ mass, the decay rate
is enhanced in SUSY by about $30 \% $ compared to the SM. 
This enhancement is difficult to disentangle 
from the non-perturbative uncertainties attendant with the SM-distributions
(shown as the shaded band in this figure).
The dilepton mass distribution  
for $B \to K^* \mu^+ \mu^-$ is more promising, as in this case the 
SUSY enhancement is around $100 \%$, see Fig.~\ref{fig:BKstsusy}, and 
this is distinguishable 
from the SM-related theoretical uncertainties (shown as the shaded band in 
this figure). The supersymmetric effects in $B \to K^* \ell^+ \ell^-$ are  
very similar to the ones worked out for the inclusive decays
$B \to X_s \ell^+ \ell^-$ \cite{goto96}, where enhancements of
($50$--$100$)\% were predicted in the low-$q^2$ branching ratios. 
The effect of $R_7$ being negative is striking in the FB 
asymmetry as shown in Fig.~\ref{fig:BKstAFBsusy}, in which the two SUGRA 
curves are plotted using Eq.~(\ref{eq:sugrar}) (for $R_7 < 0$) and
by flipping the sign of $R_7$ but keeping the magnitudes
of $R_i$ to their values given in this equation.
Summarizing for the SUGRA theories, large $\tan\beta$ solutions lead to
$\cse$ being positive, which implies that    
FB-asymmetry below the $J/\psi$-resonant region remains negative (hence, no
zero in the FB-asymmetry in this region) and one expects an 
enhancement up to a factor two in the dilepton mass distribution in $B \to 
K^* e^+ e^-$  and $B \to K^* \mu^+ \mu^-$.
\subsection{$B \to (K,K^*) \ell^+ \ell^-$ in the MIA Approach}
The minimal insertion approach aims at including all 
possible squark mixing effects in a model independent way. Choosing a 
$q,\tilde{q}$ basis where the $q-\tilde{q}-\tilde{\chi}^0$ 
and $q-\tilde{q}-\tilde{g}$
couplings are flavor diagonal, flavor changes are incorporated by a
non-diagonal mass insertion in the  
$\tilde{q}$ propagator, which can be parametrized as ($A,B=$Left, Right)
\cite{hkr86}
\begin{eqnarray}
(\delta_{ij}^{up,down})_{A,B}=
\frac{(m_{ij}^{up,down})^2_{A,B}}{m_{\tilde{q}}^2}\; ,
\end{eqnarray}
where $(m_{ij}^{up,down})^2_{A,B}$ are the off-diagonal elements of the
up(down) squark mass squared matrices that mix flavor $i$ and $j$, for
both the right- and left-handed scalars, and $m_{\tilde{q}}^2$ is the
average squark mass squared. The sfermion propagators are 
expanded in terms of the $\delta$s. The Wilson coefficients have 
the following  structure ($k=7,9,10$):
\begin{equation}
{\cal{C}}_k={\cal{C}}_k^{SM}+{\cal{C}}_k^{diag}+{\cal{C}}_k^{MIA},
\label{eq:coeffMIA}
\end{equation}
 where
${\cal{C}}^{MIA}$ is given in terms of $(\delta_{ij}^{up,down})^2_{A,B}$ up 
to two mass insertions \cite{LMSS99}, and ${\cal{C}}_k^{diag}$ being the
SUSY contribution in the basis where only flavor-diagonal contributions
are allowed.

\begin{figure}[p]
\centerline{\epsfysize=3.3in
{\epsffile{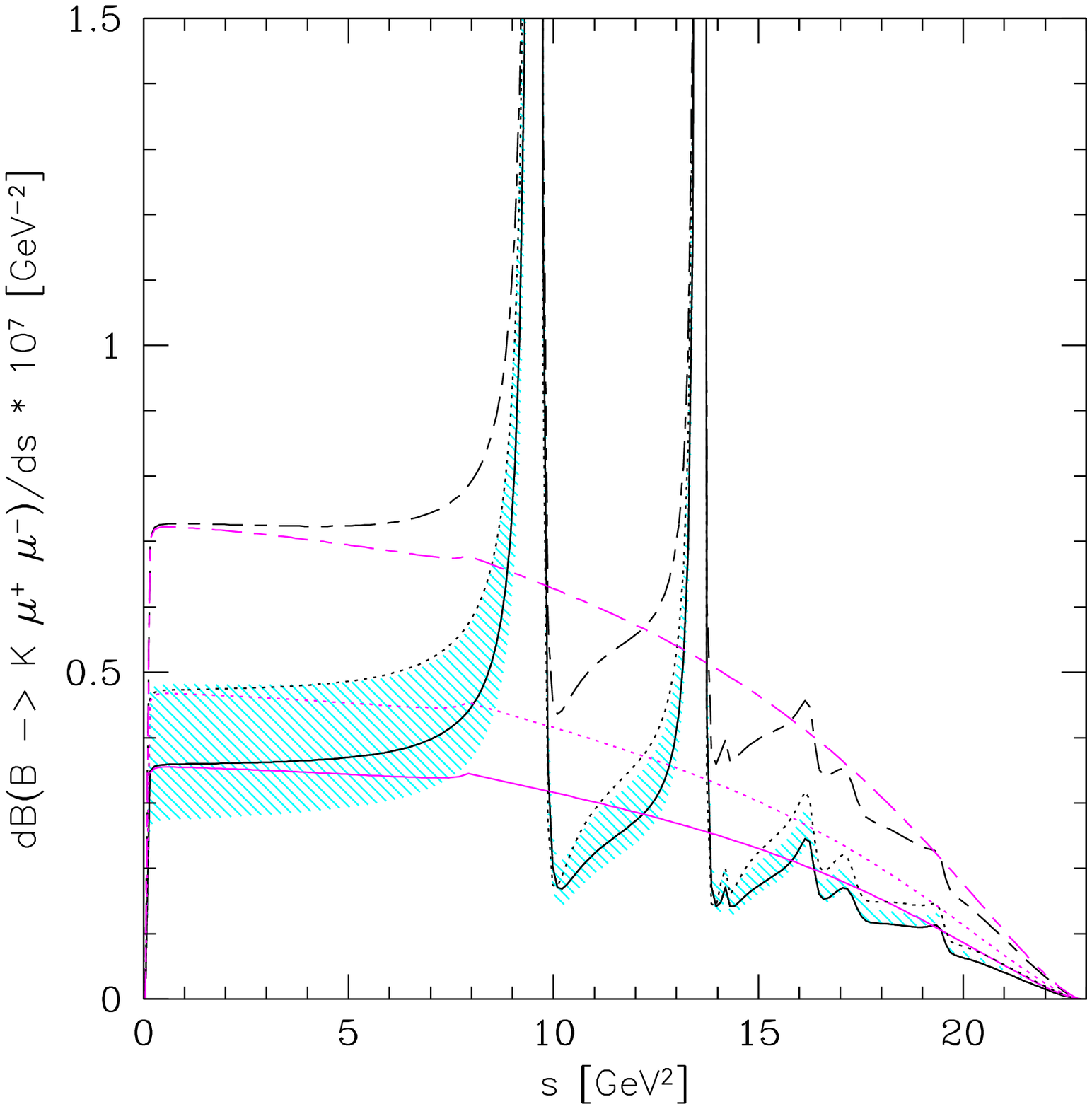}}}
\caption[]{The dilepton invariant mass distribution in
$B \to K \mu^+ \mu^-$ decays.
 The solid line represents the SM and the shaded area
depicts the form factor-related uncertainties.
The dotted line corresponds to the SUGRA model with
$R_7=-1.2,~R_9=1.03$ and $R_{10}=1$. The long-short dashed
lines correspond to the
MIA-SUSY model, given by $R_7=-0.83$, $R_9=0.92$ and $R_{10}=1.61$.
The corresponding pure SD spectra are shown in the lower part of the plot.
(From Ref.~\protect\cite{ABHH99}.)}
\label{fig:BKsusy}
$$
\epsfysize=3.3in
{\epsffile{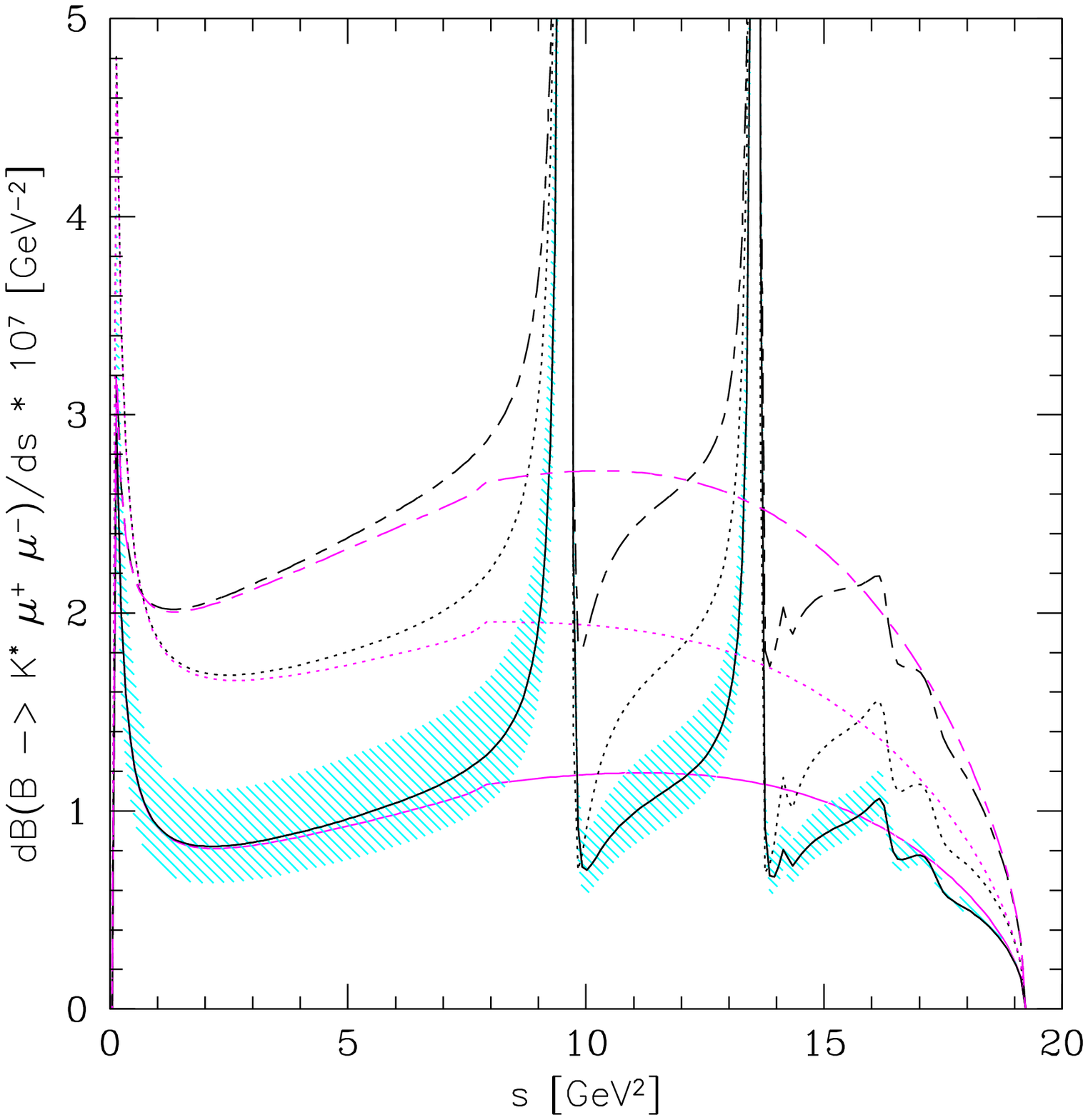}}
$$
\vspace*{-30pt}
\caption[]{The dilepton invariant mass distribution in
$B \to K^* \mu^+ \mu^-$ decays.
 Legends are the same as in
Fig.~\protect\ref{fig:BKsusy}.
(From Ref.~\protect\cite{ABHH99}.)}
\label{fig:BKstsusy}
\end{figure}

The MIA-SUSY approach has been recently used in the analysis of the decays
$B \to X_s \ell^+ \ell^-$ \cite{LMSS99}, taking into account the
present bounds on the coefficient $\cse(m_B)$ following from the
decay $B \to X_s \gamma$. The other two coefficients ${\cal{C}}_{9}^{MIA}$
and ${\cal{C}}_{10}^{MIA}$ are calculated by scanning over the allowed 
supersymmetric parameter space \cite{LMSS99}.
Illustrative examples of the dilepton invariant mass spectrum in the
decays $B \to K \mu^+ \mu^-$ and $B \to K^* \mu^+ \mu^-$ in the MIA 
approach are shown in Figs.~\ref{fig:BKsusy} and \ref{fig:BKstsusy},
respectively. They have been calculated for the following values:
\begin{equation}
R_7=\pm 0.83, ~~R_{9}=0.92, ~~~R_{10}=1.61 ~,
\label{eq:MIAplots}
\end{equation}
which are allowed by the present experimental bounds. The 
characteristic difference in this case, as compared to the SUGRA 
models, lies in the significantly enhanced value of $\ct$.   

\begin{figure}[t]
\vskip 0.0truein
\centerline{\epsfysize=3.5in
{\epsffile{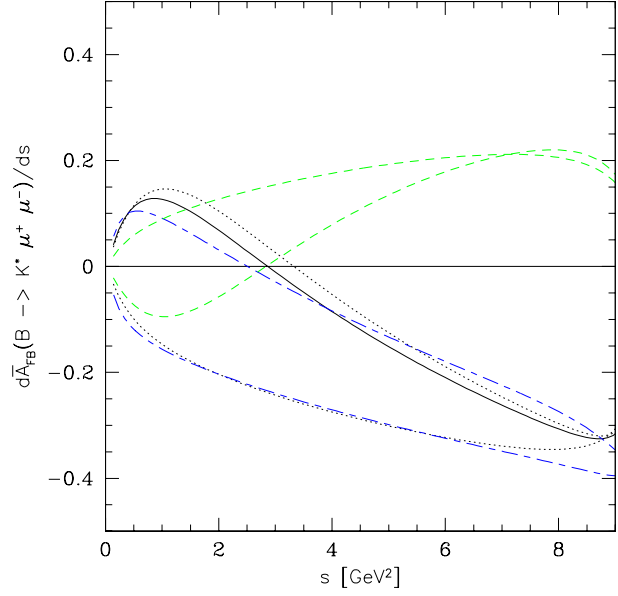}}}
\vskip 0.0truein
\caption[]{The normalized forward-backward asymmetry in
$B \to K^* \mu^+ \mu^-$ decay as a function of $s$. The
solid line denotes the SM prediction.
The dotted (long-short dashed) lines correspond to the SUGRA  
(the MIA-SUSY) model, using the parameters given in
Eq.~(\protect\ref{eq:sugrar}) (Eq.~(\protect\ref{eq:MIAplots})) with the
upper and  lower curves representing the $\cse <0$ and $\cse >0$ case,
respectively. The dashed curves indicating a positive asymmetry for
large $s$ correspond to the  MIA-SUSY models using the
parameters given in Eq.~(\ref{eq:bestdepr}).
(From Ref.~\protect\cite{ABHH99}.)}
\label{fig:BKstAFBsusy}
\end{figure}

A characteristic of the MIA approach is that 
 the sign of $\ct$ ($C_{10}^{SM}<0$) depends on the quantities
$(\delta^u_{23})_{LR}$ and $(\delta^u_{23})_{LL}$. In particular, 
in this scenario, SUSY effects could change the sign of the Wilson
coefficient $C_{10}$. This has 
no effect on the dilepton invariant mass distributions,
as they depend quadratically on $C_{10}$, 
but it would change the sign of $A_{FB}$ in $B \to K^* \ell^+ \ell^-$. To 
illustrate this, we use the following parameters
\begin{equation} 
R_7=\pm 0.83, ~~R_9=0.79, ~~R_{10}=-0.38~,
\label{eq:bestdepr}
\end{equation}
and plot the resulting normalized FB asymmetry in 
Fig.~\ref{fig:BKstAFBsusy}. The positive FB-asymmetry in $B \to K^* \ell^+
\ell^-$ (as well as in $B \to X_s \ell^+ \ell^-$ shown in 
\cite{LMSS99}) for the dilepton invariant mass below the resonant $J/\psi$ 
region is rather unique to the MIA approach.

 With $O(10^9)$   
$B\bar{B}$ events anticipated at the B-factories, and much
higher yields at the Tevatron and
LHC experiments, measurements of the differential rates and the
forward-backward asymmetry in $B \to (K,K^*) \ell^+ \ell^-$ will allow
precision tests of the SM. If we are lucky, some of these measurements  
may lead to the discovery of new physics. We illustrated the case for
supersymmetry. 


\newpage


\end{document}